\documentclass[10pt]{article}
\usepackage{latexsym,amsmath,amssymb,amsbsy,amstext,amscd,amsfonts,appendix}
\usepackage{graphics,graphicx}
\usepackage{tabularx}
\usepackage{multirow}
\usepackage{booktabs}
\usepackage{soul,color}
\usepackage{listings}
\usepackage{float}
 \usepackage{subfig,latexsym,amsmath,amssymb,amsbsy,amstext,amscd,amsfonts,appendix,color,multirow,graphicx}

\begin{document}

\title{Threshold fracture energy in solid particle erosion\\[6mm]
{\rm I. Argatov$^{\rm a,b},$
G.~Mishuris$^{\rm a}$$^{\ast}$\footnote{$^\ast$Corresponding author. Email: ggm@aber.ac.uk},\hspace{3mm}
Yu. Petrov$^{\rm c}$}\\[6mm]
{\large\rm
$^{\rm a}$ {\em{Institute of Mathematics and Physics, Aberystwyth University,
Ceredigion SY23 3BZ, Wales, UK}}};\\[6mm] 
{\large\rm
$^{\rm b}${\em{Laboratory of Friction and Wear, Institute for Problems of Mechanical Engineering of the Russian Academy of Sciences, V.O., Bolshoy pr., 61, 199178 St.~Petersburg, Russia}};}\\[6mm]
{\large\rm
$^{\rm c}${\em{Extreme States Dynamics Department, Institute for Problems of Mechanical Engineering of the Russian Academy of Sciences, V.O., Bolshoy pr., 61, 199178 St.~Petersburg, 
Russia}}}}

\maketitle

\begin{abstract}
The effect of geometrical shape of eroding absolutely rigid particles on the threshold rate of failure has been studied. The Shtaerman\,--\,Kilchevsky theory of quasi-static blunt impact, which generalizes Hertz's classical impact theory, is used for modeling the frictionless contact interaction of an axially-symmetric particle with an elastic half-space. The incubation time fracture criterion is applied for predicting surface fracture. It is shown that there exist a critical value of the particle shape parameter such that for all its lower values the fracture energy possesses a nonzero minimal value.
\end{abstract}


\bigskip

\section{Introduction}
\label{1IbSection0}

Surface fracture of materials due to solid particle erosion is encountered in a wide range of applications, including aerospace applications and degradation of components used in machinery related to mining and oil-drilling \cite{ArefiSettariAngman2005}. In recent years a number of experimental \cite{WheelerWood2009,Gnanavelu2009} and theoretical \cite{FanWang2009,JanaStack2011,Petrov2011} studies on the behavior of hard surfaces when subjected to solid particle erosion with high-impact velocities have been published. However, theoretical modeling of erosion phenomenon has still to be developed to get a deeper insight into the brittle fracture mechanisms of erosion at high impact velocities.

In analyzing dynamic strength of materials in a wide range of loading rates from quasi-static to extremely dynamic, one often encounters a contradiction between available experimental data and the classical approach for fracture characterization that is based on quasi-static consideration of the failure process
\cite{Brockenbrough1988,PetrovMorozov1994,MorozovPetrov2000}. In particular, a number of well-established experiments demonstrate that under high-rate dynamic loads, different materials are able to sustain loads substantially higher than those that cause fracture initiation in quasi-static conditions \cite{MorozovPetrov2000}. Study of this phenomenon led to the development of a concept of incubation time \cite{Petrov2004,Petrov2011}.

Most studies dealing with problems of erosion fracture \cite{BingleyOFlynn2005,HarshaBhaskar2008} develop a phenomenological approach and usually employ fracture criteria based on extrapolation of quasi-static fracture criteria to dynamic conditions. However, the temporal characteristics of the fracture process are usually not taken into consideration, though sometimes even inertia and temporal characteristics of the applied load are accounted for. In the present study, an incubation time fracture criterion \cite{Petrov2004} is used for predicting surface fracture in the erosion process.
Another assumption that is usually made is that the eroding particles are considered to be spherical. This simplifying assumption is almost always violated in reality (in particular for sand particles). The objective of this work was to develop a mathematical modeling framework for studying solid particle erosion in a wide range of particle impact velocities with varying particles shapes. One of the main results is that there exist a critical value $\lambda^*$ for the particle shape parameter $\lambda$ such that for all lower values of $\lambda$ the fracture energy possesses its nonzero minimal value.

The paper is organized as follows. In Section~\ref{1IbSection1}, we consider the quasi-static theory of blunt impact as it was developed by Kilchevsky \cite{Kilchevsky1969} based on the theory of axisymmetric frictionless contact for a polynomial indenter due to Shtaerman \cite{Shtaerman1939}. We generalize the Shtaerman\,--\,Kilchevsky theory for the case of a general non-polynomial axisymmetric indenter according to the solution obtained by Galin \cite{Galin1946}. The final form of force-displacement relationship for the blunt indenter was obtained using recently published results by Borodich and Keer
\cite{BorodichKeer2004}.
In Section~\ref{1IbSection2}, applying the approach developed by Johnson \cite{Johnson1985}, we consider the question of indentation fracture beneath a blunt indenter.
The geometry of eroding particles is described in Section~\ref{1IbSection3}.
An incubation time-based fracture criterion \cite{Petrov2004} is introduced in Section~\ref{1IbSection4}.
Prediction of the threshold fracture energy in the erosion with equiaxed superellipsoid particles is given in Section~\ref{1IbSection5}.
The developed mathematical model is illustrated in Section~\ref{1IbSection6} for the set of material properties parameters taken from the literature \cite{Petrov2011}. A discussion of the results obtained has been also outlined in Section~\ref{1IbSection6}.
Finally, in Section~\ref{1IbSection7}, we formulate our conclusion.

\section{Shtaerman\,--\,Kilchevsky theory of quasi-static blunt impact}
\label{1IbSection1}

Let us consider quasi-static indentation of an elastic half-space by a blunt indenter with the shape function
\begin{equation}
z=A r^\lambda,
\label{1Ib(1.1)}
\end{equation}
where $1<\lambda$ is a real number, $A$ is a constant having the dimension $[{\rm L}^{1-\lambda}]$ with ${\rm L}$ being the dimension of length, $r$ is a polar radius.

According to \cite{Galin1946} (see also \cite{BorodichKeer2004}), the relationship between the contact force, $P$, and the indenter displacement, $w$, can be represented as
\begin{equation}
P=k_1 w^{\frac{\lambda+1}{\lambda}},
\label{1Ib(1.2)}
\end{equation}
where
\begin{equation}
k_1=\frac{E}{1-\nu^2}A^{-\frac{1}{\lambda}}
\frac{2^{\frac{2}{\lambda}}\lambda^{\frac{\lambda-1}{\lambda}}}{\lambda+1}
\Gamma\Bigl(\frac{\lambda}{2}\Bigr)^{-\frac{2}{\lambda}}
\Gamma(\lambda)^{\frac{1}{\lambda}}.
\label{1Ib(1.3)}
\end{equation}
Here, $E$ and $\nu$ being Young's modulus and Poisson's ratio of the elastic semi-infinite body,
$\Gamma(x)$ is the Gamma function.

In the case $\lambda=2n$, where $n$ is a natural number, according to \cite{Shtaerman1939}, we will have
\begin{equation}
k_1=\frac{E}{1-\nu^2}A^{-\frac{1}{2n}}\frac{4n}{2n+1}
\biggl(\frac{(2n-1)!!}{(2n)!!}\biggr)^{\frac{1}{2n}},
\label{1Ib(1.3n)}
\end{equation}
where $(2n-1)!!=1\cdot 3\ldots (2n-1)$ and $(2n)!!=2\cdot 4\ldots (2n)$.

Let $m$ denote the mass of the indenter. Then, in view of (\ref{1Ib(1.2)}), the equation of the indenter motion can be written as follows:
\begin{equation}
m\frac{d^2 w}{dt^2}=-k_1 w^\beta.
\label{1Ib(1.4)}
\end{equation}
Here we introduced the short notation
\begin{equation}
\beta=\frac{\lambda+1}{\lambda}.
\label{1Ib(1.5)}
\end{equation}

Equation (\ref{1Ib(1.4)}) should be supplied with the initial conditions
\begin{equation}
w(0)=0,\quad \frac{d w}{dt}(0)=v_0,
\label{1Ib(1.6)}
\end{equation}
where $v_0$ is an initial velocity of the indenter.

The energy conservation law yields
\begin{equation}
\frac{m}{2}(v^2-v_0^2)=-\frac{k_1}{\beta+1}w^{\beta+1},
\label{1Ib(1.7)}
\end{equation}
where $v=dw/dt$ is the indenter velocity, and the initial conditions (\ref{1Ib(1.6)}) were taken into account.

Putting $v=0$ in Eq.~(\ref{1Ib(1.7)}), we get the maximum penetration
\begin{equation}
w_0=\biggl(\frac{(\beta+1)mv_0^2}{2k_1}\biggr)^{\frac{1}{\beta+1}}.
\label{1Ib(1.8)}
\end{equation}

The impact duration $T_0=2t_0$, where $t_0$ is the time instant at which the penetration $w(t)$ reaches its maximum $w_0$. From Eq.~(\ref{1Ib(1.7)}), it follows that
\begin{equation}
t_0=\frac{w_0}{v_0}I_\beta,
\label{1Ib(1.9)}
\end{equation}
where
\begin{equation}
I_\beta=\int_0^1\frac{dx}{\sqrt{1-x^{\beta+1}}}
=\frac{\sqrt{\pi}\Gamma\bigl(\frac{1}{\beta+1}\bigr)}{(\beta+1)\Gamma\bigl(\frac{\beta+3}{2(\beta+1)}\bigr)}.
\label{1Ib(1.10)}
\end{equation}
Note that formulas (\ref{1Ib(1.8)}) and (\ref{1Ib(1.10)}) were also derived in \cite{Borodich1989}.

Let us introduce the following dimensionless variables:
\begin{equation}
u=\frac{w}{w_0},\quad \tau=\frac{v_0}{w_0}t.
\label{1Ib(1.11)}
\end{equation}

Making use of the new variables (\ref{1Ib(1.11)}), we rewrite Eq.~(\ref{1Ib(1.4)}) and the initial conditions (\ref{1Ib(1.6)}) as follows:
\begin{equation}
\frac{d^2 u}{d\tau^2}+\frac{(\beta+1)}{2}u^\beta=0,
\label{1Ib(1.12)}
\end{equation}
\begin{equation}
u(0)=0,\quad \frac{d u}{d\tau}(0)=1.
\label{1Ib(1.13)}
\end{equation}

The unique solution of Eq.~(\ref{1Ib(1.12)}) satisfying the initial conditions (\ref{1Ib(1.13)}) will be denoted by $U(\beta;\tau)$. Thus, the dependence on time of the penetration $w(t)$ is given by
\begin{equation}
w(t)=w_0 U\Bigl(\beta;\frac{v_0}{w_0}t\Bigr),
\label{1Ib(1.14)}
\end{equation}
where $t\in[0,2t_0]$.

Integrating once Eq.~(\ref{1Ib(1.12)}) and taking into account the initial conditions (\ref{1Ib(1.13)}), we get
\begin{equation}
\biggl(\frac{d u}{d\tau}\biggr)^2+u^{\beta+1}=1.
\label{1Ib(1.15)}
\end{equation}

In the penetration stage, when $u(\tau)$ increases, Eq.~(\ref{1Ib(1.15)}) yields the following equation:
\begin{equation}
\int_0^u \frac{d\xi}{\sqrt{1-\xi^{\beta+1}}}=t.
\label{1Ib(1.16)}
\end{equation}
The above relationship represents an implicit equation for $u(t)$. Indeed, let $F(u)$ denote the left-hand side of Eq.~(\ref{1Ib(1.16)}). Then, $u(t)=F^{-1}(t)$, where $F^{-1}$ is the inverse function of $F$. Note also that $I_\beta=F(1)$ (see Eq.~(\ref{1Ib(1.10)})).

\section{Indentation fracture beneath a blunt indenter}
\label{1IbSection2}

The contact pressure beneath a blunt indenter with the shape function (\ref{1Ib(1.1)}) is given by
\begin{equation}
p(r)=\frac{(\lambda+1)}{2}p_0
\int_0^{\sqrt{1-\rho^2}}(\rho^2+\xi^2)^{\frac{\lambda-2}{2}}d\xi.
\label{1Ib(2.3)}
\end{equation}
Here, $\rho=r/a$ is the dimensionless radial coordinate, $p_0=P/(\pi a^2)$ is the mean contact pressure, $a$ is the radius of the contact area given by
\begin{equation}
a=\biggl(\frac{(1-\nu^2)P}{EA}
\frac{(\lambda+1)}{\lambda^2 2^{\lambda-1}}
\frac{\Gamma(\lambda)}{\Gamma\bigl(\frac{\lambda}{2}\bigr)^2}
\biggr)^{\frac{1}{\lambda+1}}.
\label{1Ib(2.10)}
\end{equation}

In the case $\lambda=2n$, according to \cite{Shtaerman1939}, the contact pressure can be represented in the integral form as follows:
\begin{equation}
p(r)=\frac{(2n+1)}{2}p_0 S_n\Bigl(\frac{r}{a}\Bigr)\sqrt{1-\frac{r^2}{a^2}},
\label{1Ib(2.1)}
\end{equation}
where $S_n(\rho)$ is the $n$-th Shtaerman polynomial defined as
\begin{eqnarray}
S_n(\rho) & = & \frac{(2n-2)!!}{(2n-1)!!}\biggl(
\rho^{2n-2}+\frac{1}{2}\rho^{2n-4}+\frac{3}{2\cdot 4}\rho^{2n-6}+\ldots
\nonumber \\
{} & {} & {}+ \frac{(2n-5)!!}{(2n-4)!!}\rho^2+\frac{(2n-3)!!}{(2n-2)!!}\biggr).
\label{1Ib(2.2)}
\end{eqnarray}
Correspondingly, the contact radius $a$ is related to the contact force $P$ by the equation
\begin{equation}
a=\biggl(\frac{(1-\nu^2)P}{EA}\frac{(2n+1)!!}{4n(2n)!!}\biggr)^{\frac{1}{2n+1}}.
\label{1Ib(2.10n)}
\end{equation}

The radial and circular stresses throughout the surface of the elastic half-space can be expressed in terms of the function
\begin{equation}
{\cal L}(\xi)=\int_\xi^1\frac{\eta p(\eta)\,d\eta}{(\eta^2-\xi^2)^{1/2}}
\label{1Ib(2.4)}
\end{equation}
as follows \cite{Johnson1985}:
\begin{equation}
\sigma_r(\rho)=\left\{
\begin{array}{l}
\displaystyle
-p(\rho)-\frac{2(1-2\nu)}{\pi\rho^2}\biggl(
\int_\rho^1\frac{\xi {\cal L}(\xi)\,d\xi}{(\xi^2-\rho^2)^{1/2}}
-\int_0^1{\cal L}(\xi)\,d\xi\biggr),\quad \rho\leq 1
\\
\displaystyle
\frac{2(1-2\nu)}{\pi\rho^2}\int_0^1{\cal L}(\xi)\,d\xi,\quad \rho>1,
\phantom{\Biggr)^1}
\end{array}\right.
\label{1Ib(2.5)}
\end{equation}
\begin{equation}
\sigma_\theta(\rho)=\left\{
\begin{array}{l}
\displaystyle
-2\nu p(\rho)+\frac{2(1-2\nu)}{\pi\rho^2}\biggl(
\int_\rho^1\frac{\xi {\cal L}(\xi)\,d\xi}{(\xi^2-\rho^2)^{1/2}}
-\int_0^1{\cal L}(\xi)\,d\xi\biggr)
\\
\displaystyle
\qquad {}+\frac{4}{\pi\rho}\biggl(\nu\frac{d}{d\rho}+\frac{1-\nu}{\rho}\biggr)
\int_0^\rho\frac{\xi {\cal L}(\xi)\,d\xi}{(\rho^2-\xi^2)^{1/2}},\quad \rho\leq 1
\phantom{\Biggr)^1}
\\
\displaystyle
-\frac{2(1-2\nu)}{\pi\rho^2}\int_0^1{\cal L}(\xi)\,d\xi,\quad \rho>1.
\end{array}\right.\phantom{\Biggr)^1}
\label{1Ib(2.6)}
\end{equation}

According to Eqs.~(\ref{1Ib(2.5)}) and (\ref{1Ib(2.6)}), at the edge of the contact area, we will have
\begin{equation}
\sigma_r(1)=\frac{2(1-2\nu)}{\pi}\int_0^1{\cal L}(\xi)\,d\xi.
\label{1Ib(2.7)}
\end{equation}

Now, substituting the expression (\ref{1Ib(2.4)}) into the right-hand side of Eq.~(\ref{1Ib(2.7)}) and changing the order of integration, we obtain
\begin{equation}
\sigma_r(1)=(1-2\nu)\int_0^1 \eta p(\eta)\, d\eta.
\label{1Ib(2.8)}
\end{equation}
Finally, by the definition of the contact force, we get
\begin{equation}
\sigma_r(1)=\frac{(1-2\nu)P}{2\pi a^2}.
\label{1Ib(2.9)}
\end{equation}

Note also that $\sigma_\theta(1)<0$, or, in other words, the circular stress at the edge of the contact area is compressive.
The stress field in an elastic half-space beneath an arbitrary axisymmetric indenter was studied in detail in \cite{WoirgardAudurierTromas2008}.

\section{Geometry of an axisymmetrical superellipsoid particle}
\label{1IbSection3}

A general axisymmetrical superellipsoid centered in the coordinate origin is described by the following implicit equation:
\begin{equation}
\biggl(\frac{\sqrt{x^2+y^2}}{B}\biggr)^{\lambda}+\Bigl\vert\frac{z}{C}\Bigr\vert^{\lambda}=1.
\label{1Ib(3.1)}
\end{equation}

The shape of the superellipsoid (\ref{1Ib(3.1)}) around its pole will be described by Eq.~(\ref{1Ib(1.1)}) up to the terms of higher order, if the following relation takes place between the constants $A$, $B$, and $C$:
\begin{equation}
A=\frac{C}{\lambda B^{\lambda}}.
\label{1Ib(3.2)}
\end{equation}

With $A$ and $\lambda$ being fixed constants, the shape of the superellipsoid (\ref{1Ib(3.1)}) will be defined uniquely, if we require additionally that the volume, $V$, of the superellipsoid is also fixed, that is
\begin{equation}
V=\frac{4\pi}{3\lambda}B^2 C\frac{\Gamma\bigl(\frac{2}{\lambda}\bigr)\Gamma\bigl(\frac{1}{\lambda}\bigr)
}{\Gamma\bigl(\frac{3}{\lambda}\bigr)},
\label{1Ib(3.3)}
\end{equation}
where $\Gamma(x)$ is the Gamma function.

\section{Incubation time-based fracture criterion}
\label{1IbSection4}

In this section, an incubation time fracture criterion is used to predict fracture in the case of impact interaction of a superellipsoid particle with an elastic half-space. In the simplest case, the incubation time fracture criterion is formulated as follows \cite{Petrov2004}:
\begin{equation}
\frac{1}{\tau}\int_{t-\tau}^t\sigma_r(t^\prime)\,dt^\prime=\sigma_c.
\label{1Ib(4.1)}
\end{equation}
Here, $\sigma_c$ is the tensile strength of the elastic material (this parameter is evaluated for quasi-static loading conditions), $\tau$ is the incubation time of the fracture process.

Let $\omega(t)$, $t\in[0,T_0]$, be the shape function of an impact loading pulse. Then the stress variation is given by
\begin{equation}
\sigma_r(t)=\sigma_{\max}\omega(t),
\label{1Ib(4.2)}
\end{equation}
where $\sigma_{\max}$ is the stress amplitude.

Substituting the expression (\ref{1Ib(4.2)}) into Eq.~(\ref{1Ib(4.1)}), we obtain the critical (threshold) amplitude, $\sigma_{\max}^*$, leading to fracture, i.\,e.,
\begin{equation}
\sigma_{\max}^*=\frac{\tau\sigma_c}{\displaystyle
\max_{t\in[0,T_0]}\int_{t-\tau}^t\omega(t^\prime)\,dt^\prime}.
\label{1Ib(4.3)}
\end{equation}
Here, $T_0$ is the duration of the impact process.

According to Eqs.~(\ref{1Ib(1.2)}), (\ref{1Ib(1.3)}), (\ref{1Ib(2.9)}), and (\ref{1Ib(2.10)}), we will have
\begin{equation}
\sigma_r(t)=\Pi_1(\lambda)\frac{(1-2\nu)E}{1-\nu^2}A^{\frac{1}{\lambda}}w(t)^{\frac{\lambda-1}{\lambda}},
\label{1Ib(4.4)}
\end{equation}
where we introduced the notation
\begin{equation}
\Pi_1(\lambda)=\frac{\lambda^{\frac{\lambda+1}{\lambda}}2^{\frac{2(\lambda-1)}{\lambda}}
}{2\pi(\lambda+1)}\frac{\Gamma\bigl(\frac{\lambda}{2}\bigr)^{\frac{2}{\lambda}}
}{\Gamma(\lambda)^{\frac{1}{\lambda}}}.
\label{1Ib(4.5)}
\end{equation}

On the other hand, in view of Eqs.~(\ref{1Ib(1.3)}), (\ref{1Ib(1.5)}), and (\ref{1Ib(1.14)}), we get
\begin{equation}
w(t)=w_0 U\biggl(\frac{\lambda+1}{\lambda};\frac{v_0}{w_0}t\biggr),
\label{1Ib(4.6)}
\end{equation}
\begin{equation}
w_0=\Pi_2(\lambda)\bigl(mv_0^2\bigr)^{\frac{\lambda}{2\lambda+1}}A^{\frac{1}{2\lambda+1}}
\Bigl(\frac{1-\nu^2}{E}\Bigr)^{\frac{\lambda}{2\lambda+1}},
\label{1Ib(4.7)}
\end{equation}
where
\begin{equation}
\Pi_2(\lambda)=\biggl(\frac{(2\lambda+1)(\lambda+1)}{
2^{\frac{2+\lambda}{\lambda}}
\lambda^{\frac{2\lambda-1}{\lambda}}}
\biggr)^{\frac{\lambda}{2\lambda+1}}
\biggl(\frac{\Gamma\bigl(\frac{\lambda}{2}\bigr)^2}{
\Gamma(\lambda)}\biggr)^{\frac{1}{2\lambda+1}}.
\label{1Ib(4.8)}
\end{equation}

Now, substituting (\ref{1Ib(4.6)}) and (\ref{1Ib(4.7)}) into Eq.~(\ref{1Ib(4.4)}), we obtain
\begin{equation}
\sigma_{\max}=\Pi_3(\lambda)(1-2\nu)\Bigl(\frac{E}{1-\nu^2}\Bigr)^{\frac{\lambda+2}{2\lambda+1}}
A^{\frac{3}{2\lambda+1}}\bigl(mv_0^2\bigr)^{\frac{\lambda-1}{2\lambda+1}},
\label{1Ib(4.9)}
\end{equation}
\begin{equation}
\omega(t)=U^{\frac{\lambda-1}{\lambda}}\biggl(\frac{\lambda+1}{\lambda};\frac{v_0}{w_0}t\biggr),
\label{1Ib(4.10)}
\end{equation}
\begin{equation}
\Pi_3(\lambda)=\Pi_1(\lambda)\Pi_2(\lambda)^{\frac{\lambda-1}{\lambda}}.
\label{1Ib(4.11)}
\end{equation}

In view of the notation (\ref{1Ib(1.5)}), Eq.~(\ref{1Ib(4.10)}) can be rewritten as follows:
\begin{equation}
\omega(t)=U^{2-\beta}\Bigl(\beta;\frac{v_0}{w_0}t\Bigr).
\label{1Ib(4.12)}
\end{equation}

Since the shape function (\ref{1Ib(4.12)}) of the impact loading pulse is symmetric with respect to the line $t=t_0$, where $t_0=T_0/2$, Eq.~(\ref{1Ib(4.3)}) can be transformed into the following one:
\begin{equation}
\sigma_{\max}^*=\frac{\tau\sigma_c}{\displaystyle
\int_{t_0-\tau/2}^{t_0+\tau/2}\omega(t^\prime)\,dt^\prime}.
\label{1Ib(4.14)}
\end{equation}

Let us now introduce the notation
\begin{equation}
\Upsilon(\beta,t_0)=\int_{t_0-\tau/2}^{t_0+\tau/2}
U^{2-\beta}\Bigl(\beta;\frac{v_0}{w_0}t^\prime\Bigr)\,dt^\prime.
\label{1Ib(4.13)}
\end{equation}

The evaluation of the integral (\ref{1Ib(4.13)}) depends on the relation between the impact duration $T_0=2t_0$ and the incubation time. Indeed, if $2t_0\leq\tau$, then
\begin{equation}
\Upsilon(\beta,t_0)=2\int_0^{t_0}
U^{2-\beta}\Bigl(\beta;\frac{v_0}{w_0}t^\prime\Bigr)\,dt^\prime.
\label{1Ib(4.15)}
\end{equation}
By the change of the integration variable to $s=v_0 t^\prime/w_0$, we simplify the integral (\ref{1Ib(4.15)}) as follows:
\begin{equation}
\Upsilon(\beta,t_0)=\frac{2t_0}{I_\beta}\int_0^{I_\beta}
U^{2-\beta}(\beta;s)\,ds.
\label{1Ib(4.16)}
\end{equation}

Now, if $\tau<2t_0$, then in the same way we obtain
\begin{equation}
\Upsilon(\beta,t_0)=\frac{2t_0}{I_\beta}\int_{I_\beta(1-\tau/(2t_0))}^{I_\beta}
U^{2-\beta}(\beta;s)\,ds.
\label{1Ib(4.17)}
\end{equation}
Recall that the integral $I_\beta$ is defined by formula (\ref{1Ib(1.10)}).

\section{Prediction of the threshold fracture energy}
\label{1IbSection5}

In what follows, we consider the case of an equiaxed superellipsoid particle. In other words, along with the assumptions (\ref{1Ib(3.2)}) and (\ref{1Ib(3.3)}), we put
\begin{equation}
C=B.
\label{1Ib(5.1)}
\end{equation}

Thus, according to Eqs.~(\ref{1Ib(3.2)}), (\ref{1Ib(3.3)}), and (\ref{1Ib(5.1)}), the particle's mass is given by
\begin{equation}
m=\Pi_4(\lambda)\rho_0 A^{-\frac{3}{\lambda-1}},
\label{1Ib(5.2)}
\end{equation}
where $\rho_0$ is a parameter of load intensity, having the dimension of mass density,
\begin{equation}
\Pi_4(\lambda)=\frac{4\pi}{3}\lambda^{-\frac{(\lambda+2)}{\lambda-1}}
\Gamma\Bigl(\frac{2}{\lambda}\Bigr)
\Gamma\Bigl(\frac{1}{\lambda}\Bigr)
\Gamma\Bigl(\frac{3}{\lambda}\Bigr)^{-1}.
\label{1Ib(5.3)}
\end{equation}

In view of (\ref{1Ib(5.2)}) and (\ref{1Ib(4.9)}), the maximum tensile stress generated in the impacted medium is evaluated as follows:
\begin{equation}
\sigma_{\max}=\Pi_5(\lambda)(1-2\nu)\Bigl(\frac{E}{1-\nu^2}\Bigr)^{\frac{\lambda+2}{2\lambda+1}}
\rho_0^{\frac{\lambda-1}{2\lambda+1}}v_0^{\frac{2(\lambda-1)}{2\lambda+1}}.
\label{1Ib(5.4)}
\end{equation}
Here we introduced the notation
\begin{equation}
\Pi_5(\lambda)=\Pi_3(\lambda)\Pi_4(\lambda)^{\frac{\lambda-1}{2\lambda+1}}.
\label{1Ib(5.5)}
\end{equation}

Observe that the right-hand side of Eq.~(\ref{1Ib(5.4)}) does not depend on the geometrical parameter $A$.

Further, according to Eqs.~(\ref{1Ib(1.9)}), (\ref{1Ib(1.8)}), and (\ref{1Ib(5.2)}), we obtain the penetration duration $t_0$ in the form
\begin{equation}
t_0=\Pi_6(\lambda)\Bigl(\frac{1-\nu^2}{E}\Bigr)^{\frac{\lambda}{2\lambda+1}}
\rho_0^{\frac{\lambda}{2\lambda+1}}A^{-\frac{1}{\lambda-1}}v_0^{-\frac{1}{2\lambda+1}},
\label{1Ib(5.6)}
\end{equation}
where
\begin{equation}
\Pi_6(\lambda)=I_\beta \Pi_2(\lambda)\Pi_4(\lambda)^{\frac{\lambda}{2\lambda+1}}.
\label{1Ib(5.7)}
\end{equation}

Following \cite{Petrov2011}, we extract the quantities $A$ and $v_0$ from the system (\ref{1Ib(5.4)}), (\ref{1Ib(5.6)}). First of all, Eq.~(\ref{1Ib(5.4)}) immediately yields
\begin{equation}
v_0=\biggl(\frac{\sigma_{\max}}{(1-2\nu)\Pi_5(\lambda)}\biggr)^{\frac{2\lambda+1}{2(\lambda-1)}}
\Bigl(\frac{1-\nu^2}{E}\Bigr)^{\frac{\lambda+2}{2(\lambda-1)}}\rho_0^{-\frac{1}{2}}.
\label{1Ib(5.8)}
\end{equation}

Second, substituting the expression (\ref{1Ib(5.8)}) for $v_0$ into the right-hand side of Eq.~(\ref{1Ib(5.6)}), we obtain after some algebra the following equation:
\begin{equation}
A^{-\frac{1}{\lambda-1}}=\frac{t_0}{\Pi_6(\lambda)}
\biggl(\frac{\sigma_{\max}}{(1-2\nu)\Pi_5(\lambda)}\biggr)^{\frac{1}{2(\lambda-1)}}
\Bigl(\frac{E}{1-\nu^2}\Bigr)^{\frac{\lambda-2}{2(\lambda-1)}}\rho_0^{-\frac{1}{2}}.
\label{1Ib(5.9)}
\end{equation}

Evaluating the initial kinetic energy, $\varepsilon_0$, of the superellipsoid particle, one can estimate the energy required to create fracture in the impacted medium as follows:
\begin{equation}
\varepsilon_0=\alpha_\lambda\frac{\displaystyle
t_0^3(\sigma_{\max})^{\frac{4\lambda+5}{2(\lambda-1)}}}{\displaystyle
\rho_0^{\frac{3}{2}}E^{\frac{10-\lambda}{2(\lambda-1)}}}.
\label{1Ib(5.10)}
\end{equation}
Here, $\alpha_\lambda$ is a dimensionless coefficient given by
\begin{equation}
\alpha_\lambda=\frac{\Pi_4(\lambda)}{2\Pi_6(\lambda)^3}\frac{\displaystyle
(1-\nu^2)^{\frac{10-\lambda}{2(\lambda-1)}}}{\displaystyle
\bigl((1-2\nu)\Pi_5(\lambda)\bigr)^{\frac{4\lambda+5}{2(\lambda-1)}}}.
\label{1Ib(5.11)}
\end{equation}
Note that $\alpha_\lambda$ depends on Poisson's ratio (this is not indicated in notation for simplicity).

\section{Numerical results and discussions}
\label{1IbSection6}

As shown in a number publications \cite{PetrovMorozovSmirnov2003,Petrov2004,TulerButcher1968,OuDuanHuang2010,Vignjevic-Panov2012}, the criterion (\ref{1Ib(4.1)}) can be successfully used for predicting the fracture initiation and structural transformations in brittle solids under dynamic loading. For slow loading rates (and, correspondingly, times to fracture that are essentially bigger than $\tau$), the condition (\ref{1Ib(4.1)}) for crack initiation gives the same predictions as the classic critical stress criterion. For high loading rates (and, correspondingly, times to fracture comparable with $\tau$), all the variety of effects experimentally observed in dynamic fracture experiments \cite{MorozovPetrov2000,PetrovMorozovSmirnov2003} can be explained using (\ref{1Ib(4.1)}), both qualitatively and quantitatively. It should be emphasized that the criterion itself gives correct predictions in a wide range of loading rates, and it is not necessary to pay a special attention to the particular time scale of the impact problem.

In our numerical calculations, the properties of the impacted material are taken to be equal to the properties of gabbro-diabase \cite{Petrov2011} as follows:
$E=6{.}2\times 10^9$ $\rm N/m^2$,
$\nu=0{.}26$,
$\sigma_c=44{.}04\times 10^6$ $\rm N/m^2$,
$\tau=44$ $\mu{\rm s}$. The mass density of eroding particles is $\rho_0=2400$ $\rm kg/m^3$.

\begin{figure}[h!]
\hspace{4mm}\includegraphics [scale=0.34]{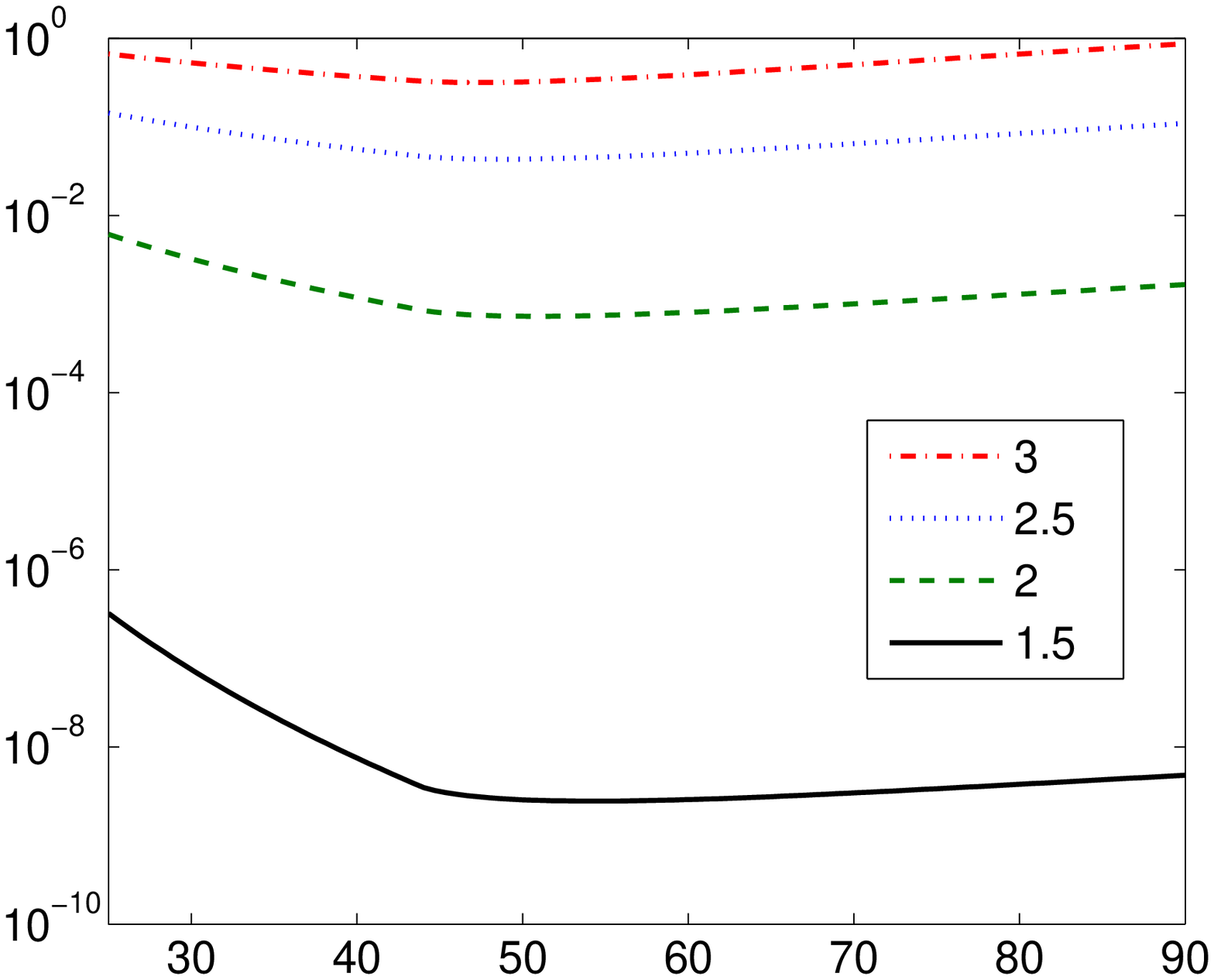}\hspace{-2mm}
\includegraphics [scale=0.34]{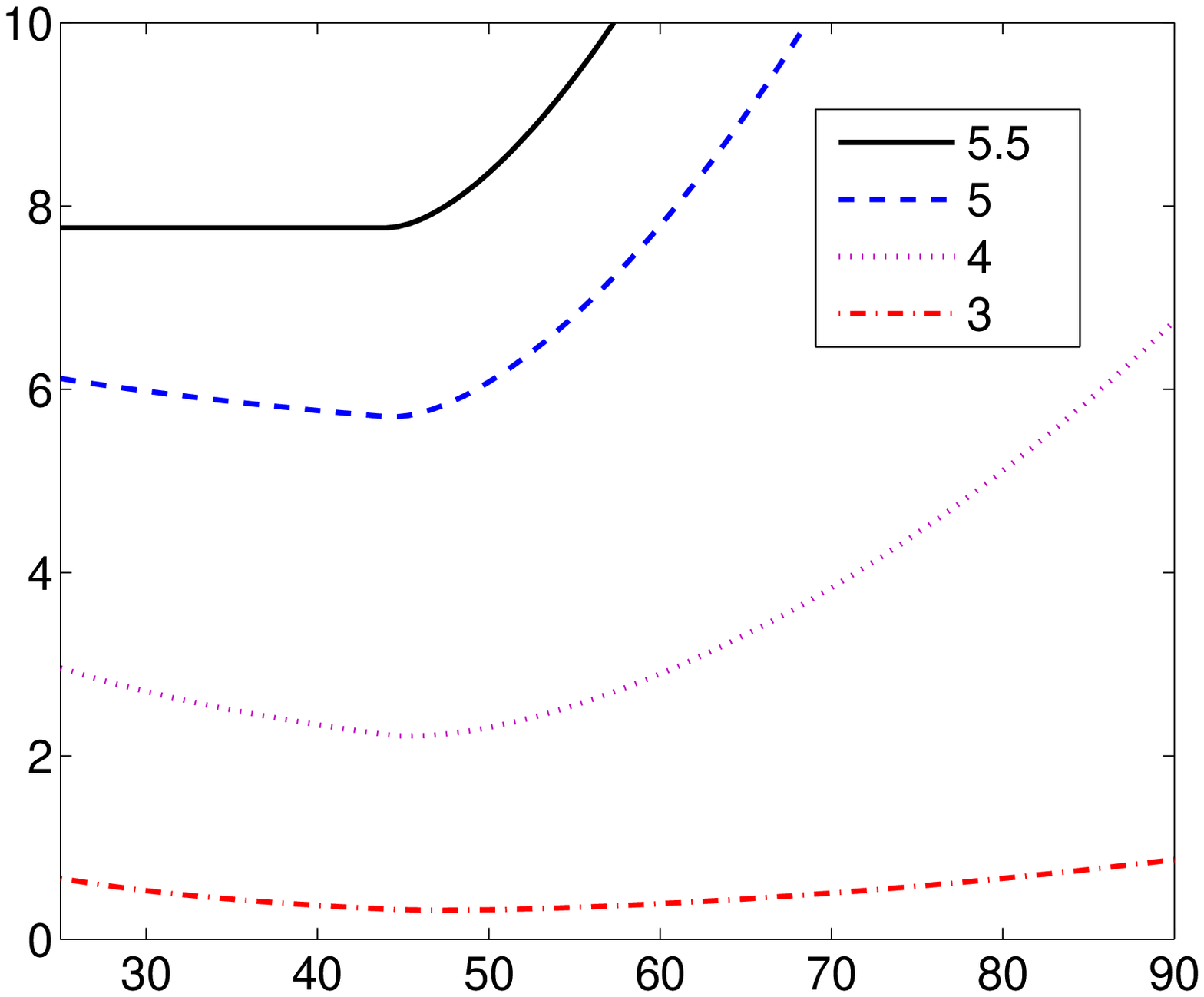}
\begin{picture}(0,0)(-6,0)
\put(0,140){a)}
\put(195,140){b)}
\put(-4,94){$\varepsilon_0$}
\put(98,5){$2t_0$} \put(290,5){$2t_0$}
\end{picture}
\caption{The initial kinetic energy, $\varepsilon_0$, as a function of the impulse length, $2t_0$,
for different shapes of the  blunt indenter determined by the value
of the parameter $\lambda$ indicated in the legend sections (Fig.~\ref{fig:1}a and Fig.~\ref{fig:1}b).  For all $\lambda<5.5$ the energy
possesses its nonzero minimal value. In the {\it critical} case $\lambda=5.5$ the minimal energy corresponds to the horizontal part of the graph.}
\label{fig:1}
\end{figure}

First of all, observe (see Fig.~\ref{fig:1}a) that with decreasing the shape parameter $\lambda$, the fracture energy $\varepsilon_0$ decreases too. This dependency is so strong that a logarithmic scale is required to show the trend. Recall that the case $\lambda=2$  corresponds to spherical particles, while with decreasing $\lambda$, the contact surface shape approaches a cone (with the limit value $\lambda=1$).
It is clear that infinitesimally small fracture energies are required for a sharp indenter, which produces a singular contact pressure distribution. By the way, the same can be said of a cylindrical indenter with a sharp edge. Note also that under the assumption of elastic impact the fracture energy coincides with the kinetic energy of eroding particles.

{\bf Remark.}
It is interesting to observe that for all $\lambda<5.5$ the fracture energy possesses a nonzero minimal value. In other words, there exists an energetically optimal mode of impact erosion such that the energy input for fracture is minimal. From the practical point of view, this effect can be used, for example, for optimization of the polishing process so that the energy cost of the process is minimized \cite{Volkov-Petrov2009}.
After determining the critical value $t_0^*$, for which the fracture energy $\epsilon_0$ takes its minimum, Eq.~(\ref{1Ib(5.6)}) allows one to determine the corresponding value $v_0^*$ for the eroding flow velocity.
In view of Eq.~(\ref{1Ib(5.2)}), we will have
$$
t_0^*=I_\beta\Pi_2(\lambda)\Bigl(\frac{1-\nu^2}{E}\Bigr)^{\frac{\lambda}{2\lambda+1}}
m^{\frac{\lambda}{2\lambda+1}}A^{\frac{1}{2\lambda+1}}(v_0^*)^{-\frac{1}{2\lambda+1}}.
$$
Because $t_0^*$ depends on $\lambda$, the optimal value $v_0^*$ will also depend on this shape parameter as well as on the material parameters $E$ and $\nu$, mass of eroding particles $m$ and on the geometrical parameter $A$. In view of Eqs.~(\ref{1Ib(3.2)}) and (\ref{1Ib(5.1)}),
we have the relation $A=\lambda^{-1}C^{1-\lambda}$, where $C$ is the particle semi-axis.
It should be noticed, of course, that the eroding particles do usually have a wide range of size and shape parameters and this circumstance should be taken into account. Also, friction does play a role, especially, for oblique impacts \cite{Argatov-Petrov2009}. In the case of normal impact incidence, as a first approximation, friction effects can be usually neglected.

However, it is more interesting to find that there is the so-called {\it critical} value $\lambda^*=5{.}5$ of the shape parameter, when the minimal energy corresponds to the horizontal part of the graph (see Fig.~\ref{fig:1}b). This means that for $\lambda=\lambda^*$, the fracture energy remains the same for small values of the duration of loading impulse $2t_0$ comparable with the incubation time $\tau$.

\begin{figure}[h!]
\centering
\includegraphics [scale=0.34]{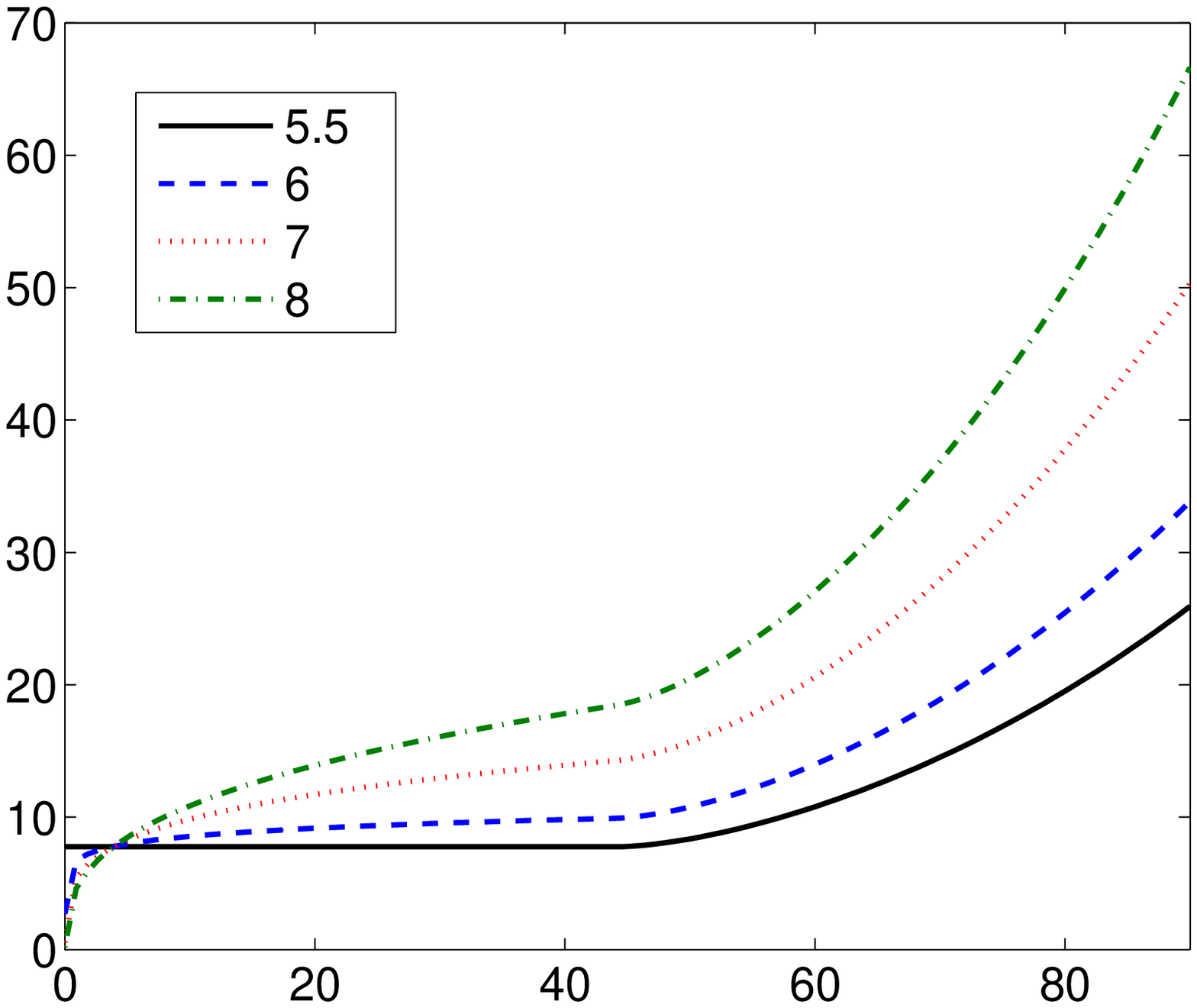}
\begin{picture}(0,0)(110,0)
\put(-85,80){$\varepsilon_0$}
\put(3,-4){$2t_0$}
\end{picture}
\vspace*{3mm}
\caption{The initial kinetic energy, $\varepsilon_0$, as a function of the impulse length, $2t_0$,
for the shapes of the  blunt indenter determined by the parameter $\lambda$ larger than its {\it critical} value $\lambda^*$, that is $\lambda>5{.}5$.
There is no minimum for the energy (the shorter impulse the lower energy level is accessible).}
\label{fig:2}
\end{figure}

Fig.~\ref{fig:2} illustrates the influence of the impact duration $2t_0$ on the fracture energy $\varepsilon_0$ for blunt particles with the super-critical shape of the contacting surface. The absence of the minimum can be explained by the fact that the contacting surface with increasing its bluntness leads to the contact stress concentration.


\begin{figure}[h!]
\hspace{4mm}\includegraphics [scale=0.34]{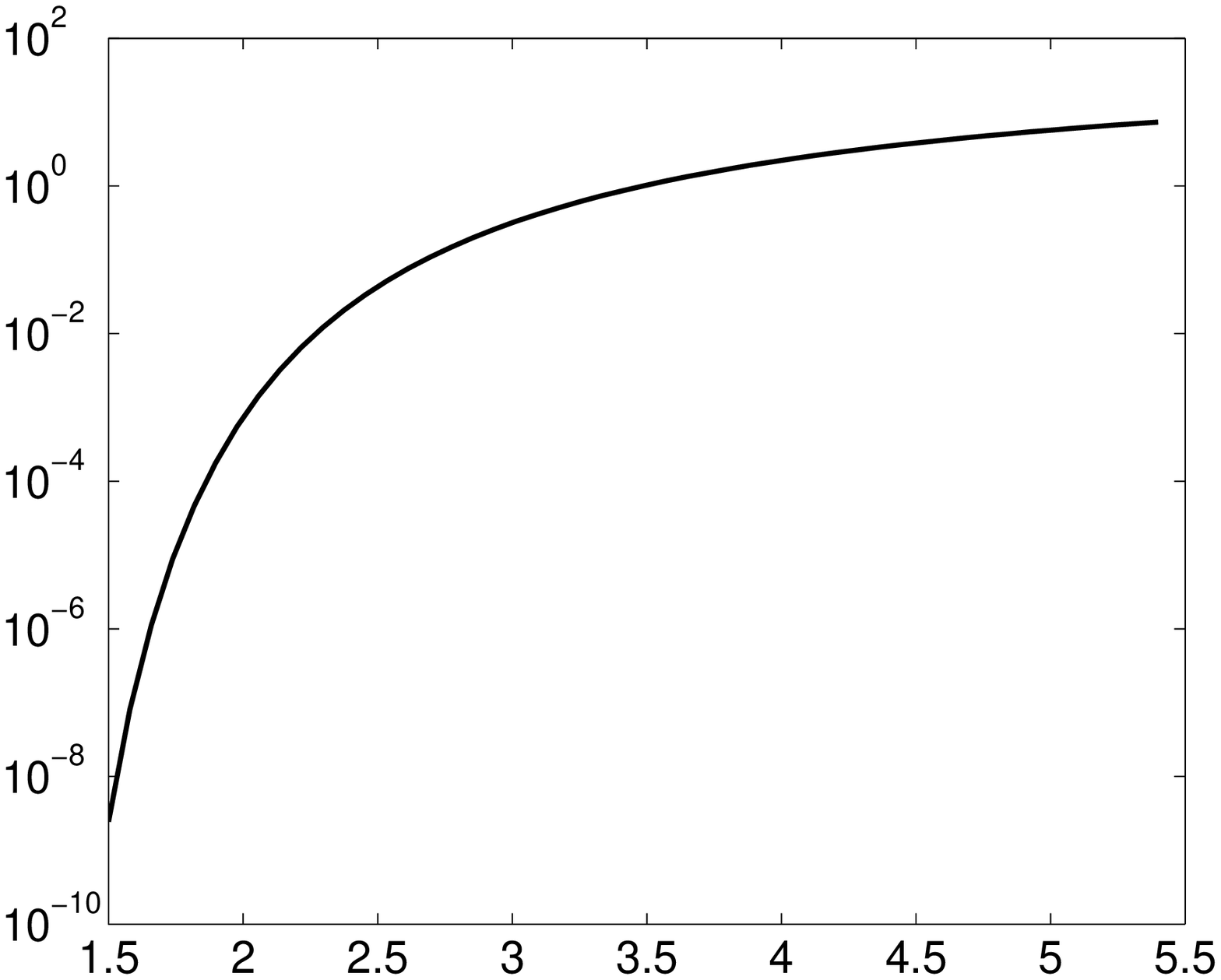}\hspace{1mm}
\includegraphics [scale=0.34]{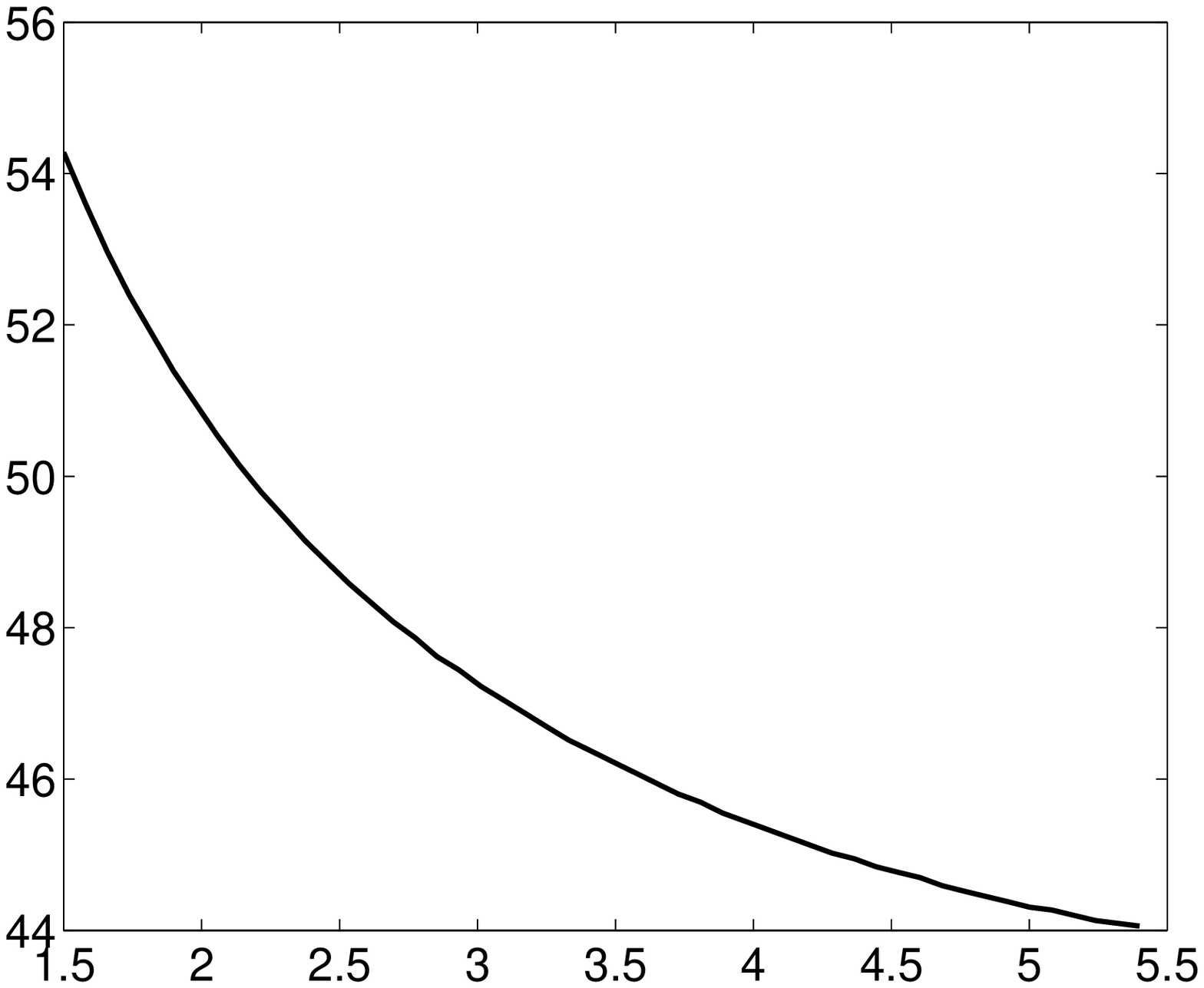}
\begin{picture}(0,0)(-6,0)
\put(0,140){a)}
\put(200,140){b)}
\put(-7,95){$\varepsilon_{\rm min}$}
\put(195,95){$\tau_{\rm min}$}
\put(100,5){$\lambda$} \put(295,5){$\lambda$}
\end{picture}
\caption{Fig.~\ref{fig:4} a): The minimum of the initial kinetic energy, $\varepsilon_0$, in the logarithmic scale as a function of the parameter $\lambda$ determining the shape of the  blunt indenter. Fig.~\ref{fig:4} b): The length of the impulse, $\tau_{\rm min}$, corresponding to the minimal value of the initial kinetic energy, $\varepsilon_{\rm min}$.}
\label{fig:4}
\end{figure}

Finally, Figs.~\ref{fig:4}a and \ref{fig:4}b present the variation of the minimum $\varepsilon_{\rm min}$ of the fracture energy $\varepsilon_0$ and the corresponding impact duration $\tau_{\rm min}$ with variation of the shape parameter $\lambda$. It is seen that with $\lambda$ approaching its critical value $\lambda^*$, the optimal impact duration $\tau_{\rm min}$ approaches the incubation time $\tau$.

\section{Conclusion}
\label{1IbSection7}
 	
In the present paper, the effect of geometrical shape of eroding absolutely rigid particles on the threshold rate of failure has been studied. The eroding particles are assumed in the form of axisymmetrical equiaxed superellipsoid, and the case of normal impact is considered. The dynamic contact interaction between an axially-symmetric particle with an elastic half-space has been described in the framework of the Shtaerman\,--\,Kilchevsky theory of quasi-static blunt impact.
Stress field created at the contact surface as a result of the impact interaction is estimated using Galin's solution of the contact problem for a blunt indenter.
For predicting surface fracture beneath a frictionless blunt indenter, we employ the incubation time fracture criterion.

The effect of geometric shape of eroding particles on the threshold rate of failure was very recently considered in \cite{VolkovGorbushinPetrov2012},
where the following three special cases were studied: spherical
\cite{Volkov-Petrov2009,Petrov2011},
cylindrical \cite{Smirnov2007},
and paraboloidal shape with $\lambda=4$.
In all papers mentioned above, the particle motion function was assumed according to the sine law, while in the present analysis the particle equation of motion is solved implicitly. However, the main advantage of the employed approach concerns the developed mathematical modeling framework for studying the geometry effect in the whole range of the particle shape parameter $\lambda$.
In the case of spherical particles, the present analysis reduces to the theory developed in \cite{Volkov-Petrov2009,Petrov2011}
with the only difference that the actual particle motion slightly differs from the sine law.
However, in the other two cases, the developed approach differs also by the use of the equiaxed superellipsoid shape for eroding particles.

Based on the present analysis, the kinetic energy necessary for a rigid particle to fracture the impacted surface is analyzed. Threshold particle velocity (and, hence, threshold kinetic energy) corresponding to the initiation moment of surface fracture in the elastic half-space is determined.
As the main results of the present study it is shown that the value of the fracture energy does significantly depend on the load duration and has a marked minimum in the so-called subcritical case when $\lambda<\lambda^*$. It is also shown that $\lambda>\lambda^*$ the fracture energy achives its zero minimal value with decreasing impact duration.
Existence of energetically optimal modes of dynamic impact is claimed. In particular, it is demonstrated that the energy input for fracture can be optimized so that the energy cost of the process is minimized. The effect itself turns out to be dependent on the bluntness of the particle's contacting surface.

\section*{Acknowledgements}

I.A. and Yu.P. were supported by the Russian Foundation for Basic Research (project Nos. 10-08-00966-a and 11-01-00491-a); G.M. gratefully acknowledges the support from the European Union Seventh Framework Programme Marie Curie Programme (project No. PIAP-GA-284544-PARM-2).

\end{document}